\documentclass[twocolumn,aps,amssymb]{revtex4}
\usepackage{graphicx}
\usepackage{epsfig}
\newcommand{\be}{\begin{equation}}
\newcommand{\ee}{\end{equation}}
\newcommand{\bea}{\begin{eqnarray}}
\newcommand{\eea}{\end{eqnarray}}

\newcommand{\bc}{\begin{center}}
\newcommand{\ec}{\end{center}}
\usepackage[colorlinks,bookmarks=false,citecolor=blue,linkcolor=red,urlcolor=blue]{hyperref}
\usepackage{graphicx,natbib,ulem}
\usepackage{epstopdf}
\usepackage{color}

\begin{document}
\title{Finite temperature ordering of dilute graphene antiferromagnets}
\author{Thomas Fabritius${}^1$, Nicolas Laflorencie${}^2$ and Stefan Wessel${}^1$}
\affiliation{${}^1$Institut f\"ur Theoretische Physik III,
Universit\"at Stuttgart, Pfaffenwaldring 57, 70550 Stuttgart, Germany}
\affiliation{${}^2$Laboratoire de Physique des Solides,
Universit\'e Paris-Sud, UMR-8502 CNRS, 91405 Orsay, France}

\begin{abstract}
We employ large-scale quantum Monte Carlo simulations to study the magnetic
ordering transition among dilute magnetic moments randomly localized on the graphene honeycomb lattice, induced by long-ranged RKKY interactions
at low charge carrier concentration. In this regime the effective exchange interactions are ferromagnetic within each 
sublattice,
and antiferromagnetic between opposite sublattices, with an overall cubic decay of the interaction strength with 
the separation between the moments. We verify explicitly, that this commensurability leads to antiferromagnetic
order among the magnetic  moments below a finite transition temperature in this two-dimensional system.
Furthermore, the  ordering temperature shows a crossover in its power-law 
scaling with the moments' dilution from a low- to a high-concentration regime. 

\end{abstract}
\pacs{75.10.Jm, 75.30.Hx, 75.40.Mg}
\maketitle
\section{Introduction}
\begin{figure}[t]
\begin{center}
\includegraphics[width=7cm]{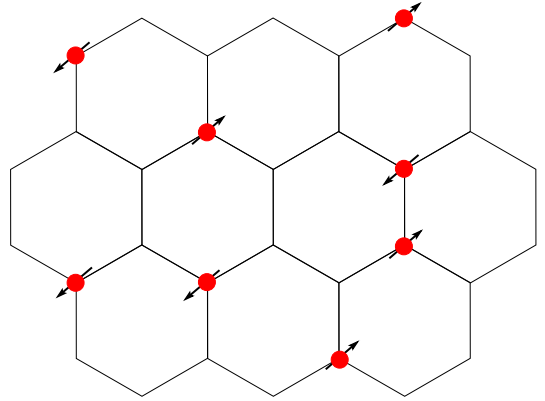}
\caption{(Color online) Magnetic moments localized on a honeycomb lattice. Due to the commensurate nature of the long-ranged exchange interaction, 
the moments order antiferromagnetically below a finite transition temperature.}
\label{fig_setup}
\end{center}
\end{figure}

Since the discovery of the single-layer hexagonal carbon sheets of graphene, its fascinating physical properties have
stimulated a lot of research on the
static and transport properties of fermions on the two-dimensional hexagonal lattice~\cite{geim07a, castroneto09a}. 
Soon after its discovery, the role of disorder effects on graphene has opened up new research horizons~\cite{peres06,aleiner06}. In particular, disorder induced 
localized states~\cite{pereira06} and magnetism~\cite{vozmediano05,kumazaki07,uchoa08,palacios08} have attracted a lot of interest. For instance, it was
found that
graphene's  electronic properties give rise  to the efficient formation of magnetic moments from adatoms located on the 
graphene surface~\cite{uchoa08}{, or simply from defects~\cite{vozmediano05,kumazaki07,palacios08}.}
Motivated by such studies, 
we here consider a situation such as shown in Fig.~\ref{fig_setup}, 
where {local moments are induced in a graphene sheet by for instance adatoms or defects}, each single magnetic moment being associated with a particular lattice site 
of the underlying {lattice}~\cite{brey07,saremi07}. 
At low carrier concentration, 
the low density of states near the Dirac points suppresses the Kondo  effect and allows for the formation of magnetically ordered states by the  interaction between the localized moments and the conduction electrons~\cite{brey07,saremi07,sengupta08}. Namely, 
they induce  {{long-ranged}} Ruderman-Kittel-Kasuya-Yosida (RKKY) exchange interactions between the localized magnetic moments, described by the Hamiltonian
\begin{equation}
 {{H}}= \sum_{i,j} J_{ij}^{{\rm RKKY}} \:\mathbf{S}_i\cdot \mathbf{S}_j.
 \label{eq:H}
\end{equation}
The effective interaction $J_{ij}^{{\rm RKKY}}$, mediated by itinerant electrons, strongly depends on the electronic properties at the Fermi energy. While in typical metals an oscillating coupling at $2k_F$ is expected ($k_F$ being the Fermi wave vector), decaying as $1/r^{2}$ in two dimensions ($r$ being the relative distance between impurities), 
the semi-metallic properties of graphene lead to a different behavior~\cite{vozmediano05,brey07,saremi07}. Indeed, the absence of an extended Fermi surface leads to a cancellation of the $1/r^2$ term and leaves the next term decaying as $1/r^3$ without any $2k_F$ oscillations. Furthermore, it was revealed in Refs.~\cite{brey07,saremi07}  that pair-wise interactions are ferromagnetic (antiferromagnetic) among the same (different) sublattice on the bipartite honeycomb lattice of graphene.

This means for the exchange couplings in Eq.~(\ref{eq:H}), that
\be
J_{ij}^{{\rm RKKY}}=\epsilon_{ij}J({|\mathbf{r}_i-\mathbf{r}_j}|),\quad J(r)=\frac{J_{}}{r^3},
\label{eq:JRKKY}
\ee
with $J_{}>0$ and $\epsilon_{ij}=-1$ $(+1)$ if $i$ and $j$ 
belong to the same (different) sublattice~\cite{brey07,saremi07}. Here, $\mathbf{r}_i$ denotes the  (random) position of the $i$-th magnetic moment 
$\mathbf{S}_i$ 
on the honeycomb lattice, which we consider to be spin-$\frac{1}{2}$ quantum spins, in order to explore the extreme quantum limit. 
{Larger spin values of the moments will not qualitatively change the results}. The commensurate nature of the RKKY interactions was linked to the 
bipartiteness of the underlying  lattice geometry~\cite{saremi07}. 
In a more recent work~\cite{bunder09}, this general result was called into question for graphene nanoribbons,
due to the presence of zero-mode contributions. In bulk  graphene however, these  corrections vanish in the thermodynamic limit~\cite{bunder09}, thus recovering the commensurate form of the interactions in Eq.~(\ref{eq:JRKKY}).
In the following, 
we  focus  on the most basic model that contains the main features of such exchange interactions (i.e. their commensurate, long-ranged nature) between the magnetic moments,
and leave for discussion at the end of the paper several aspects relevant to graphene, such as doping effects, a finite extension of the localized moments, and structural defects.

The remainder of this paper is organized as follows: In the following section, we review some general results concerning long-range order in low-dimensional systems. Then, we present our numerical approach in Sec. III. The results of our simulations are discussed in Sec. IV and Sec. V. Concluding remarks are made in Sec. VI, while an appendix provides  details about the relevant length and energy scales on the diluted honeycomb lattice. In the appendix, we  furthermore introduce the notion of an effective coordination number for diluted magnetic moments, that will be convenient for the discussion in Sec. V. 

\section{long-range order in 2D}

Starting from the effective exchange interactions of Eq.~(\ref{eq:JRKKY}) in Eq.~(\ref{eq:H}), we explore its consequences for the finite-temperature ordering transition between magnetic moments on the honeycomb lattice.
The stability of  long-range magnetic order in $d\le 2$ systems with power-law decaying interactions has been the subject of a large number of theoretical studies in the past~\cite{mermin66,fisher72,brezin76,cardy81,haldane88,shastry88,nakano94,nakano95,vassiliev01,bruno01,luijten02,yusuf04,laflorencie05,beach07}. Regarding the Heisenberg model with $1/r^{\alpha}$ interactions, the seminal paper of Mermin and Wagner~\cite{mermin66}, proving the absence of finite-$T$ spontaneous order if $\alpha>d+2$, was recently reconsidered by Bruno~\cite{bruno01}, who gave stronger conditions, notably on the appearance of ferromagnetism for oscillatory interactions. For instance, an interaction of the form $\cos(k_0 r)/r^{\alpha}$ ($k_0\neq 0$) cannot lead to finite-temperature ferromagnetism if $\alpha>5/2$. For the case under study here, Bruno's result implies the absence of finite-temperature antiferromagnetic order for $\alpha>2d$ which does not deviate from Mermin-Wagner's results in $d=2$.

Early renormalization group calculations on classical O($n$) models with $1/r^{d+\sigma}$ couplings predicted~\cite{fisher72,brezin76} a $\sigma$-dependent criticality with a finite ordering temperature for $\sigma<d$: $T_c\propto (d-\sigma)/(n-1)$ when $\sigma\to d$~\cite{brezin76}. For instance, while for $\sigma<d/2$ exponents take exact "classical" values ($\eta=2-\sigma$, $\nu=1/\sigma$, $\gamma=1$), our case $\sigma=1$ lies on the boundary of this regime where logarithmic corrections are expected~\cite{fisher72}. 
In particular, the correlation length exponent turns out to be $\nu=1^{+}$, which fulfills the Harris criterion~\cite{harris74,vojta06} $\nu>2/d$. From such a statement we expect on general grounds that clean and disordered systems with power-law interactions display similar critical behaviors if $\sigma\le 1$.
We indeed find, that the model in Eq.~(\ref{eq:H}) exhibits at finite temperature a phase transition to an antiferromagnetically ordered N\'eel state, both in case of a fully covered lattice and also for the case of diluted magnetic moments, with apparently similar critical exponents.

In the following, we analyze in detail the dependence of the ordering temperature on the concentration $p$ of the magnetic moments. While in the realistic parameter regime the magnetic moments will be dilute, i.e. $p\ll 1$, we  consider for completeness the whole range up to (and including) the case of full coverage $p=1$, where a magnetic moment resides on every lattice site.  For the full coverage case, we also performed simulations for  an underlying square lattice, in order to explore  more generally the magnetic ordering transition in quantum antiferromagnets with long-range interactions in two dimensions. While in previous works~\cite{nakano94,nakano95,vassiliev01}, the case of solely ferromagnetic interactions on a square lattice geometry has been analyzed, the current case and the effects of dilution have not been considered thus far.

\section{Methods}

\begin{figure}[t]
\begin{center}
\includegraphics[width=8.5cm]{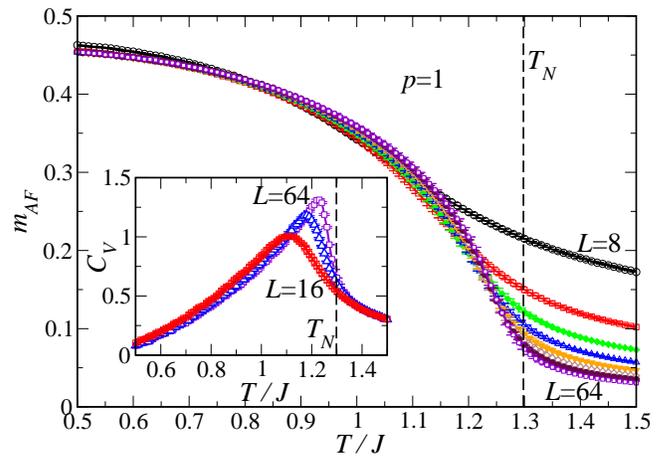}
\caption{(Color online) 
Temperature dependence of the staggered magnetization $m_{AF}$ for different system sizes $L=8, 16, 32, 40, 48, 56$, and 
$64$ for the occupation density $p=1$ (full coverage). The inset shows the temperature dependence of the specific heat $C_V$ for system sizes 
$L=16, 48$, and $64$. The dashed lines indicate the position of the transition temperature as estimated from a finite size scaling analysis of 
the Binder parameter. 
}
\label{fig_m_AF_gamma1}
\end{center}
\end{figure}
For our study, we employed large-scale quantum Monte Carlo simulations based on the stochastic series expansion representation~\cite{sandvik99b}, with an improved diagonal update scheme adapted to systems with long-ranged interactions~\cite{sandvik03}. In addition, we used Walker's method of alias~\cite{walker77} in order to speed up the algorithm~\cite{fukui09}. We performed simulations on finite systems with $N_l=2L^2$ lattice sites and linear system sizes ranging up to $L=192$, depending on the concentration $p$ of the magnetic moments. 
For $p<1$, we performed statistical averages over independent realizations 
of the moments' distribution in a canonical ensemble, such that  no sample-to-sample fluctuations in the total number of moments $N=p\times N_l$ result. Typically, we performed disorder averages over several thousand realizations, and verified that the calculated observables followed Gaussian distributions, as expected.  We always employed period boundary conditions, and 
used the minimum image convention for the $1/r^3$-decaying exchange constants.  
For each choice of $p$ and $L$, we measured the staggered magnetization, obtained using the standard operator
\begin{equation}
 \mathbf{S}_{AF}=\frac{1}{N}\sum_i^N \epsilon_i \mathbf{S}_i, 
\end{equation}
after performing the disorder averaging as 
\begin{equation}
m_{AF}=\sqrt{[\langle \mathbf{S}_{AF}^2\rangle]_{av}}.
\end{equation}
Here, $\epsilon_i=\pm 1$, depending on the sublattice to which spin $i$ belongs on the honeycomb lattice, $\langle...\rangle$ denotes the  QMC statistical mechanics expectation value for each realization of disorder, and $[...]_{av}$ the final disorder averaging. 
We also calculated the Binder parameter
\begin{equation}
Q=[\langle \mathbf{S}_{AF}^4\rangle]_{av}/[\langle \mathbf{S}_{AF}^2\rangle]_{av}^2.
\end{equation}
We then used a finite size scaling analysis to extract the critical temperature and  exponents from the finite size data of $m_{AF}$ and $Q$. 
More details about the finite size scaling analysis are provided below. 

\begin{figure}[t]
\begin{center}
\includegraphics[width=8.5cm]{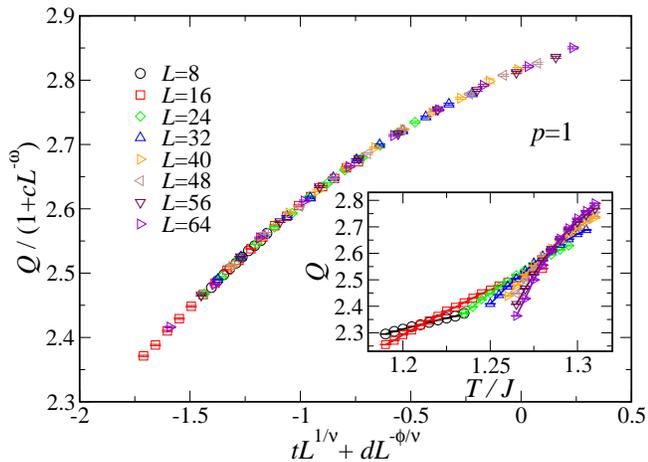}
\caption{(Color online)
Data collapse of the Binder parameters for the full case $p=1$ in a finite size scaling analysis. Here,
$t=(T-T_N)/T_N$ denotes the reduced temperature. The inset shows the   Binder parameters $Q$ for different system sizes  in the vicinity of the 
consecutive crossing points.
}
\label{fig_binder-gamma1} 
\end{center}
\end{figure}

{{\section{Full coverage}}}
As a useful starting point in the absence of any disorder, we consider firstly the full coverage case $p=1$.
In Fig.~\ref{fig_m_AF_gamma1}, we present the QMC data for the temperature dependence of the staggered magnetization for various system sizes. This data shows, that a finite temperature ordering transition takes place near $T\sim 1.3J$. This is also evident from the behavior of the specific heat $C_V$, shown in the inset of Fig.~\ref{fig_m_AF_gamma1}. It also exhibits  pronounced finite size effects, that need to be accounted for in order to extract the  transition temperature. 

For this purpose, we calculated the Binder parameter $Q$ inside the transition region, for which the finite size data is shown in 
the inset of Fig.~\ref{fig_binder-gamma1}. The strongly moving crossing points in the data for consecutive system sizes again indicate significant finite size effects, that need to be taken into account in the further analysis. 
We thus employed a finite size scaling analysis including leading corrections to scaling~\cite{beach05}, used 
in several precision studies on  critical properties in quantum spin systems~\cite{wang06,wenzel08a,wenzel09a}.

It is based on  a scaling ansatz for the Binder parameter
\begin{equation}
Q(t,L)=(1+c L^{-\omega})g(tL^{1/\nu}+d L^{-\phi/\nu}),
\label{scalingansatz}
\end{equation}
with the reduced temperature $t=(T-T_N)/T_N$, and the scaling function $g$. From this analysis, the critical exponent $\nu$ is  also obtained. Furthermore, $\omega, \phi, c$, and $d$ describe the leading corrections to scaling, which are necessary in order to fit the QMC data obtained here for a system with long-ranged interactions on the limited system sizes available to our numerical study. 
Following Ref.~\onlinecite{wang06}, we represent $g$ up to forth-order in a Taylor expansion ($g(x)=g_0+g_1 x + g_2 x^2 + g_3 x^3 + g_4 x^4$), and use  bootstrapping  in combination with a standard Levenberg-Marquardt nonlinear optimization algorithm to perform the minimization procedure and fit the numerical data to the above scaling from.

In Fig.~\ref{fig_binder-gamma1}, we show the resulting data collapse for the case of $p=1$. 
\begin{figure}[t]
\begin{center}
\includegraphics[width=8.8cm]{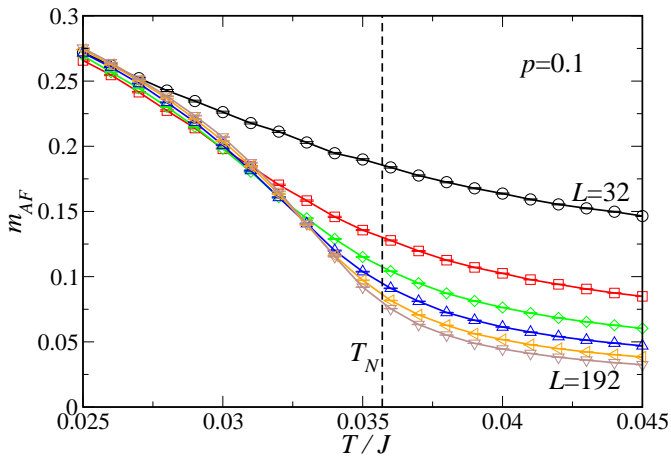}
\caption{(Color online) 
Temperature dependence of the staggered magnetization $m_{AF}$ for different system sizes $L=32, 64, 96, 128, 160$, and 
$192$ for the occupation density $p=0.1$. The dashed line indicates the position of the transition temperature as estimated from a finite size 
scaling analysis of the Binder parameters.
}
\label{fig_m_AF_gamma01}
\end{center}
\end{figure}
We find that the QMC data can be fitted  well to the above scaling form, leading to an estimate of the transition temperature of $T_N/J=1.298\pm 0.001$ 
{
with three significant digits. While the other fitting parameters are less constrained by the finite-size data (see below) -- as observed also in the above-mentioned high-precision studies of short-range interacting quantum spin systems~\cite{wang06,wenzel08a,wenzel09a} --
we  obtain from the finite-size analysis robust estimates of $T_N$, which is the quantity we are mainly interested in for this study;
in particular, since we will analyse its dependence on the dilution $p$ in the following section.
}
Furthermore, we obtain 
{
an estimate for the correlation length critical exponent
} $\nu=1.04\pm 0.02$. This value 
is in good agreement with the predicted value $\nu=1$ from the renormalization group approach~\cite{fisher72}, the small deviations from this prediction 
being attributed to
logarithmic corrections
in the dependence of the correlation length on the reduced temperature~\cite{fisher72}. However, given the restricted range of system sizes available to our QMC study, we are not in a position, to accurately account for these additional corrections. 
{
From the same finite-size data, we also obtain estimates for $\omega=0.4\pm0.1$ and $\phi=0.6\pm0.1$, which  are less constrained within the bootstrapping analysis. Similar as for $\nu$, we  expect residual finite-size effects also on these values due to the logarithmic corrections. These values are about a factor of two smaller than the values given e.g. in Ref.~\onlinecite{wang06} for the case of the quantum phase transition in bilayer Heisenberg models, where however the expected universality class is that of  the three-dimensional Heisenberg transition instead of the mean-field behavior expected here. 
From the fitting procedure, we obtain non-zero values for both prefactors of the subleading finite-size corrections, $d=-2\pm0.6$, and $c=-0.11\pm0.06$. This exhibits the necessity of including the subleading finite-size corrections to the leading scaling behavior in Eq.~(\ref{scalingansatz}).
The Taylor expansion coefficients of the scaling function $g$
can be estimated from the bootstrapping analyis as 
$g_0=2.93\pm 0.04$, 
$g_1=0.18\pm 0.04$,
$g_2=-0.1\pm 0.01$,
$g_3=-0.001\pm 0.008$,
$g_4=0.009\pm 0.003$.
The last two coefficients remain more unconstrained than the other fitting parameters. This indicates that $g$ could also be represented well by a second order polynomial within the considered region close to $T_N$ with coefficients similar to those given above.  
}

We performed an analysis for $p=1$ also for the case of an underlying square lattice, and obtained the transition temperature in that case to be $T_N/J=1.855\pm0.02$, which is in fact close to the values obtained from QMC simulations and from using a Green's function decoupling for the fully ferromagnetic case~\cite{vassiliev01, nakano95}. Furthermore, for the square lattice, we obtain
an estimate of $\nu=1.13\pm 0.07$, which  within the error bars agrees with the result for the honeycomb lattice, but deviates more  from the mean-field value. From the finite size scaling of $m_{AF}\propto L^{-\beta/\nu}$ at $T_N$, we extract the ratio $\beta/\nu=0.52\pm0.04$, which within error bars agrees  with the value $\beta/\nu=1/2$ from  renormalization group calculations~\cite{fisher72}. 
{
The estimates for
$\omega=0.4\pm0.1$ and $\phi=0.6\pm0.1$, that we obtain for the square lattice, also agree 
within the error bars  with the result for the honeycomb lattice. Again, this is expected, as the ordering transitions on both lattices belong to the same universality class.
} 
\\
\\

\begin{figure}[t]
\begin{center}
\includegraphics[width=8.5cm]{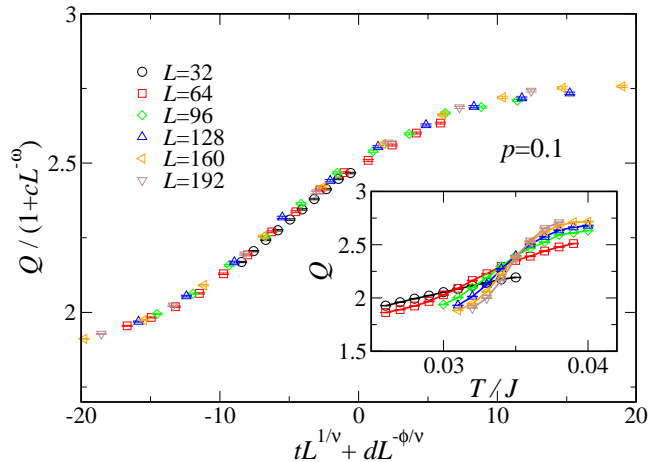}
\caption{(Color online) 
Data collapse of the Binder parameters for $p=0.1$ in a finite size scaling analysis. Here,
$t=(T-T_N)/T_N$ denotes the reduced temperature. The inset shows the   Binder parameter 
$Q$ for different system sizes taken in the vicinity of the 
consecutive crossing points. 
}
\label{fig_binder-gamma01} 
\end{center}
\end{figure}

%
{{\section{Randomly Diluted Moments}}}
After having considered the full coverage limit, we now turn to the case of diluted magnetic moments, $p<1$. Also in this case  we do obtain a finite temperature ordering transition. For example, the QMC data for the staggered magnetization at $p=0.1$ is shown in Fig.~\ref{fig_m_AF_gamma01}. 
The corresponding  data for the Binder parameter is shown in the inset of Fig.~\ref{fig_binder-gamma01}. 
Performing
the same finite size scaling analysis as before, we  estimate the N\'eel temperature as $T_N/J=0.0357\pm 0.0006$ for $p=0.1$. 
The corresponding data collapse of the Binder parameter is shown
in the main panel of Fig.~\ref{fig_binder-gamma01}.
{
The estimate for the correlation length critial exponent $\nu=1.0\pm0.02$ appears somewhat closer to the mean-field value, while the results for $\omega=0.4\pm0.1$ and $\phi=0.6\pm0.1$  agree well with the above values at $p=1$.
} 

\begin{figure}[t]
\begin{center}
\includegraphics[width=8.5cm]{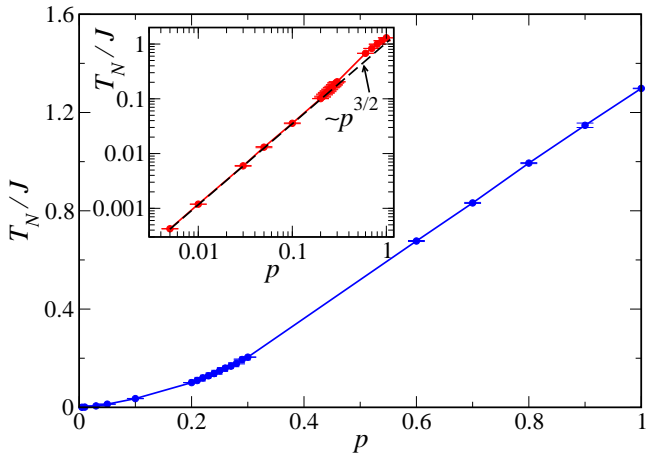}
\caption{(Color online) 
Magnetic transition temperature $T_N$ as a function of the occupation density $p$. The inset shows
the same data on a log-log plot. A power-law dependence proportional to $p^{3/2}$ for $p\lesssim0.2$ is indicated by the dashed line. 
}
\label{fig_tc}
\end{center}
\end{figure}

Proceeding in the same way for various values of $p$, we eventually obtain the dilution dependence of the N\'eel temperature shown in Fig.~\ref{fig_tc}, which summarizes the main results from our numerical study. 
{
Concerning the estimates for the exponents $\nu$, $\omega$ and $\phi$, we cannot observe, within  statistical errors, any systematic changes with $p$ from their values in the clean limit, which appears consistent with the discussion in Sec.~II. 
}

On the other hand, the transition temperature shows a strong dependence on $p$ that we now analyse further. 
In the range $0.3\lesssim p <1$, this dependence is almost perfectly linear. An extrapolation of the linear suppression would exclude finite-temperature magnetic order below $p\sim 0.18$. However, we find the low-$p$ behavior of $T_N$ to deviate from this linear behavior. In fact, as seen from the inset of  Fig.~\ref{fig_tc}, which shows the same data on a log-log plot, below $p\sim 0.2$, $T_N$ exhibits an algebraic increase with $p$, scaling as
\begin{equation}
T_N\propto p^{3/2},\quad p\lesssim 0.2.
\end{equation}
In the following, we discuss the relevant energy scales behind the different behavior of $T_N$ at high and low concentration of the magnetic moments:

On a two-dimensional lattice,  dilute randomly distributed
magnetic moments are separated by {{an average}} distance that scales as $\langle r\rangle \propto p^{-1/2}$ (on the honeycomb lattice $\langle r\rangle = p^{-1/2}/2$, cf. the appendix) which defines a {\it{typical}} coupling strength  $J_{\rm typ}=J(\langle r\rangle)\propto p^{3/2}$.
In the low-$p$ regime, {{we thus find that the}} N\'eel temperature scales with the characteristic energy scale set by $J_{\rm typ}$.
At higher concentrations, the scaling of $T_N$ with $p$ becomes more mean-field-like, namely directly proportional to the mean-field average coupling (see Eq.~(\ref{eq:MF}) in the appendix) $J_{\rm avg}^{\rm MF}\propto p$.
This leads to the linear behavior in $T_N$ observed at higher values of $p$.
In this regime, 
the average nearest-neighbor distance between the magnetic moments is $\langle r\rangle \sim 1$, and does not vary much as a function of $p$. Its main effect is the reduction of exchange paths, as the number of bonds that each moment is associated with reduces linearly with $p$. 
The 
crossover results near $p\sim 0.25$, corresponding to a 
concentration regime beyond which the average distance becomes $\langle r\rangle \sim 1$, as shown in the appendix.

The behavior of $T_N$ can be qualitatively understood also with the help of a $p$-depended effective coordination number $Z_{\rm eff}(p)$, as defined in Eq.~(\ref{eq:Zeff}), which displays two distinct regimes: For large dilution $p\ll 1$ (i.e. for $\langle r \rangle \gg 1$), $Z_{\rm eff}\sim 1$, and the natural energy scale for the magnetic ordering is set by the coupling value at the average distance (i.e. $J_{\rm typ}$), because each moment has only a few neighbors to couple with and thus $T_N\sim J_{\rm typ}$. Increasing $p$, once $Z_{\rm eff}$ becomes significantly larger than one (i.e. once $\langle r \rangle \sim 1$) the relevant energy scale which controls the ordering of the moments will be controlled by the average $J_{\rm avg}^{\rm MF}$, directly proportional to $p$. As shown in the appendix, the crossover between these two regimes takes place near $p\sim 0.25$. It is interesting to compare such a concentration to the percolation threshold of the 2D honeycomb lattice $p^*\sim 0.697$~\cite{suding99} where nearest-neighbor interacting quantum spins lose  long-range magnetic order in the ground state~\cite{castro06}. The absence of any feature at $p^*$ in the present study is in fact consistent with  the sizeable value of the effective coordination number $Z_{\rm eff}(p^*)\sim 6.5$.\\
\\

\section{Discussion and conclusions}
Motivated by recent results on the properties of RKKY interactions between localized magnetic moments  on graphene~\cite{brey07,saremi07},
we performed a systematic study of the finite-temperature ordering transition of dilute spin-1/2 magnetic moments on the honeycomb lattice, induced by a commensurate long-ranged exchange interaction. We found that in the low dilution regime, {{where the effective coordination number is close to unity (i.e. the average separation $\langle r\rangle \gg 1$)}}, the N\'eel temperature scales {{with the typical coupling $J_{\rm typ}\propto p^{3/2}$}}. For larger occupations, the behavior crosses over to a {{mean-field-like}} linear reduction of the N\'eel temperature from its value in the full coverage case.
We also presented estimates for the critical exponents $\beta$ and $\nu$, which  within statistical errors are consistent with the prediction from previous renormalization group calculations for the ferromagnetic classical $O(3)$ model, given that additional logarithmic corrections are expected~\cite{fisher72}. 
In our analysis, we considered the extreme quantum limit of $S=1/2$ magnetic moments. However, the physical picture will not change   except that the N\'eel temperature will scale with $S(S+1)$ for higher quantum spins.
For the future, it will be interesting to explore the critical properties of such diluted quantum magnets with long-ranged exchange interactions in more detail, also considering other decay rates of the exchange interactions. This would require the consideration of significantly larger lattices. Our main focus here was on the diluted case, relevant to the physical situation in graphene. 

In the case of graphene, the RKKY coupling, controlled by the ratio between Coulomb repulsion $U$ and band-width $W$, is $J\sim U^2/W\sim 1$ eV~\cite{vozmediano05} which for a moderate concentration $p\sim 10^{-2}$ would give a critical temperature $\sim 10 $K. Of course this estimate is based on a very simple model of localized point-like magnetic impurities.
A more realistic description should be able to incorporate (i) the spatial extension $\xi$ of the defects, (ii) the holes/electrons doping effects, (iii) lattice distortions (ripples for instance). Regarding (i), a finite area $\xi^2$ for a defect is expected to move the crossover concentration $p\sim 0.25$ above which MF behavior $T_N\sim p$ occurs towards a lower value $p\sim 0.25/\xi^2$. (ii) As already discussed in Ref.~\cite{vozmediano05}, holes/electrons doping  shifts the Fermi energy, thus leading to a finite Fermi wave vector $k_F\sim\sqrt {n_{c}}$ ($n_c$ being the carriers concentration). RKKY interactions will oscillate with a wave length $\lambda_F\sim 1/\sqrt{n_c}$, while the average distance between moments is $\langle r\rangle \sim 1/\sqrt{p}$. Therefore the above analysis, which ignores $2k_F$ oscillating terms is expected to be valid provided $n_c\ll p$. Alternatively, one expects the N\'eel order to be destroyed upon carrier doping in graphene sheets. 
{For instance, using an electric field to control the Fermi level would render it possible to induce a transition from the N\'eel ordered regime for 
$k_F\ll \sqrt{p}$ onto a more complex regime at $k_F\sim \sqrt{p}$ where competing interactions, 
i.e. magnetic frustration, associated with random dilution are expected to provide all the ingredients to achieve spin-glass physics.}
We note that in the commensurate case at half-filling, the ferromagnetic and the antiferromagnetic exchange interactions actually have different prefactors~\cite{saremi07}. However, this does not lead to any frustration, and hence  including these prefactors will not destroy the finite temperature antiferromagnetic state.
(iii) With respect to lattice distortions, it would  be interesting to account for the characteristic ripples in the graphene structure~\cite{geim07a, castroneto09a} and explore its consequences on the magnetic order, given the long-ranged nature of the exchange interactions.
This would extend a recent study that considered this interplay between structural and magnetic properties within an effective Ising model with exponentially suppressed exchange interactions on the order of several lattice spacings~\cite{rappoport09}. 

Two directions appear feasible to experimentally probe for the two-dimensional magnetism in graphene at finite 
temperatures: using magnetic adatoms like Mn for instance, or extrinsic defects
~\footnote{Note that intrinsic defects are present in graphene, with a concentration $p_{\rm int}\lesssim 10^{-4}$.} 
that could be created by irradiation.
In addition to randomly distributed moments, it will be interesting to explore the situation considered in Ref.~\onlinecite{uchoa08}, where the 
magnetic 
moments are placed 
using STM techniques onto specific lattice sites, and to examine the magnetic states induced by the RKKY interactions. 
For such studies, the effects of frustration  could lead to  exotic magnetic phases, the study of which is however  
beyond the scope of the quantum Monte Carlo approach, due to the infamous sign-problem~\cite{troyer05a}. 
In that respect, future experiments on graphene might even be employed as a quantum simulator for such magnetic clusters. 

\begin{acknowledgements}
We thank M. Barbosa da Silva Neto, H. Bouchiat, J.-N. Fuchs, and M.-O. Goerbig
for helpful discussions, and in particular F. Alet for suggesting to us this investigation. 
Furthermore,  we
acknowledge the allocation of CPU time on the HLRS Stuttgart 
and NIC J\"ulich supercomputers. 
\end{acknowledgements}
{{
\appendix
\section{Energy scales on the diluted honeycomb lattice}
In order to gain insight into the role played by various energy scales, we performed a numerical analysis on a $L=5000$ diluted system, introducing a fraction $p$ of magnetic moments randomly on the honeycomb lattice. The nearest-neighbor distance (i.e. the distance from a randomly chosen moment to the closest other moment) obeys a probability distribution (see the inset of Fig.~\ref{fig_avgJ}),
that at low concentrations $p$ is very well described by 
\be
P(r)=2\pi r p \exp(-\pi p r^2),
\ee
thus resulting in an average nearest-neighbor distance $\langle r\rangle =p^{-1/2}/2$, shown in  Fig.~\ref{fig_avgr}. This leads to a \textit{typical} coupling strength $J_{\rm typ}=J(\langle r\rangle)\propto p^{3/2}$. It is interesting to compare this to the {\it{average nearest-neighbor}} coupling $J_{\rm avg}^{\rm nn}$, defined as
\be
J_{\rm avg}^{\rm nn}=\int J(r) P(r){\rm d}r,
\ee
which, at low concentration $p\ll 1$, turns out to be (i) much larger than $J_{\rm typ}$ and (ii) a linear function of $p$. On the other hand, the \textit{mean-field} \textit{average} coupling 
\be
J_{\rm avg}^{\rm MF}=\frac{1}{N}\sum_{i,{j\neq i}}J({|\mathbf{r}_i-\mathbf{r}_j}|)
\label{eq:MF}
\ee
compares well to $J_{\rm avg}^{\rm nn}$ at low doping. But while $J_{\rm avg}^{\rm MF}$ remains  linear ($J_{\rm avg}^{\rm MF}= 2\pi a_{\rm hex}\zeta(3) p$~\footnote{$\zeta(3)=\sum_{p=1}^{\infty}p^{-3}$ is the Riemann-zeta function, and $a_{\rm hex}$ is a geometric factor for the hexagonal lattice that we estimate to be about $1.2$.}) as $p$ increases, $J_{\rm avg}^{\rm nn}$ approaches $J_{\rm typ}$ for larger $p$. 
This $p$-dependence of the different energy scales is shown in Fig.~\ref{fig_avgJ}.

The effective coordination number, defined as
\be
Z_{\rm eff}=J_{\rm avg}^{\rm MF}/J_{\rm avg}^{\rm nn},
\label{eq:Zeff}
\ee
clearly
traces these two different regimes. For $\langle r\rangle \gg 1$ (i.e. beyond $p\ll 1$), the system is highly diluted and $Z_{\rm eff}\sim 1$ increases only slightly with $p$, whereas once $\langle r\rangle \sim 1$ (i.e. beyond $p\sim 0.25$), the effective number of magnetic neighbors increases much more rapidly, proportional to $p$. This difference in behavior directly follows from Fig.~\ref{fig_avgr}.
\begin{figure}[t]
\begin{center}
\includegraphics[width=6.5cm,angle=90]{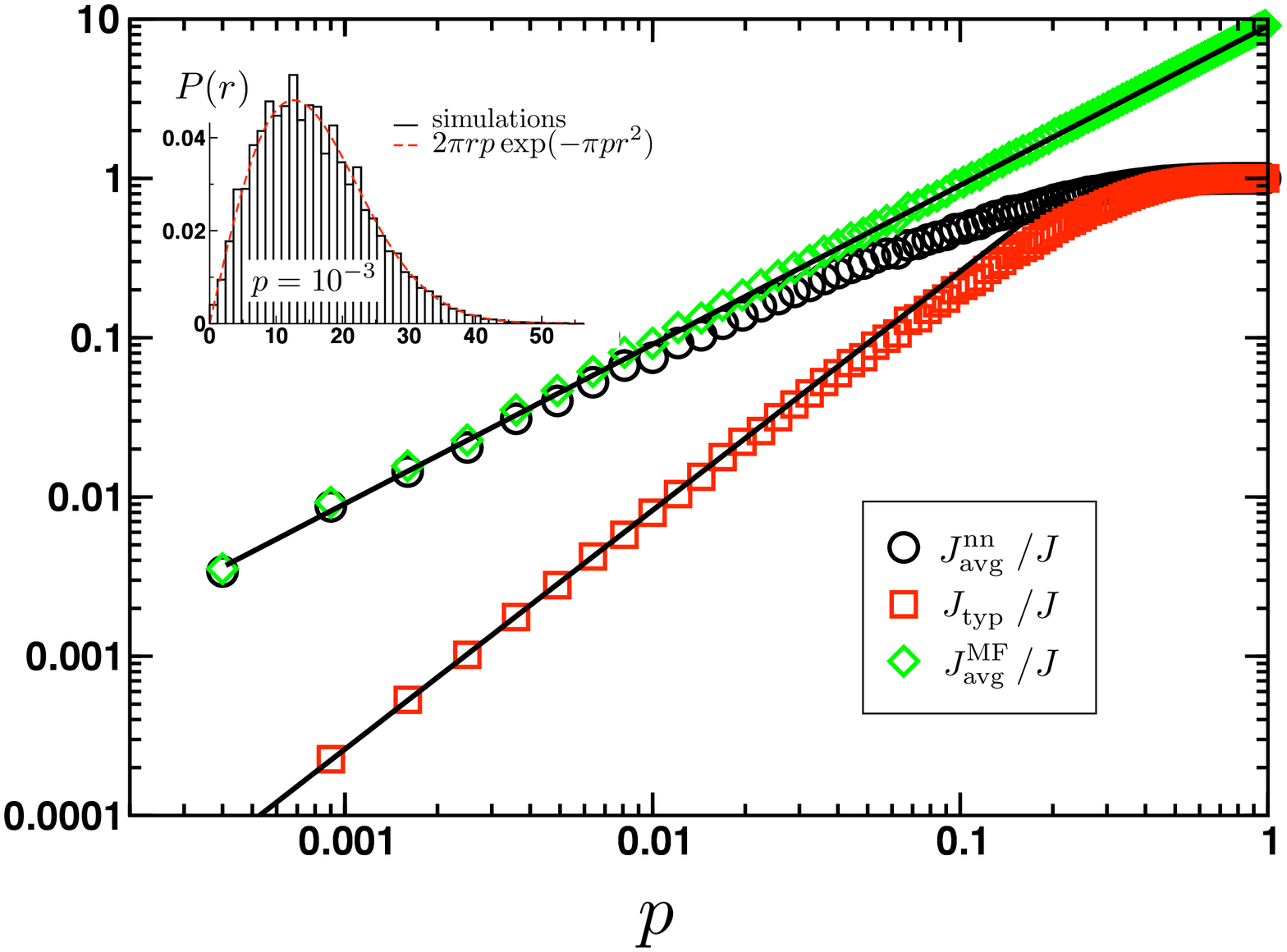}
\caption{{(Color online) Dependence of the various energy scales  on the
occupation density $p$. The inset shows the histogram of the nearest-neighbor distance obtained from simulations 
of a $L=5000$ system, along with the analytic formula for $P(r)$.
}}
\label{fig_avgJ}
\end{center}
\end{figure}
\begin{figure}[b]
\begin{center}
\includegraphics[width=6.5cm,angle=90]{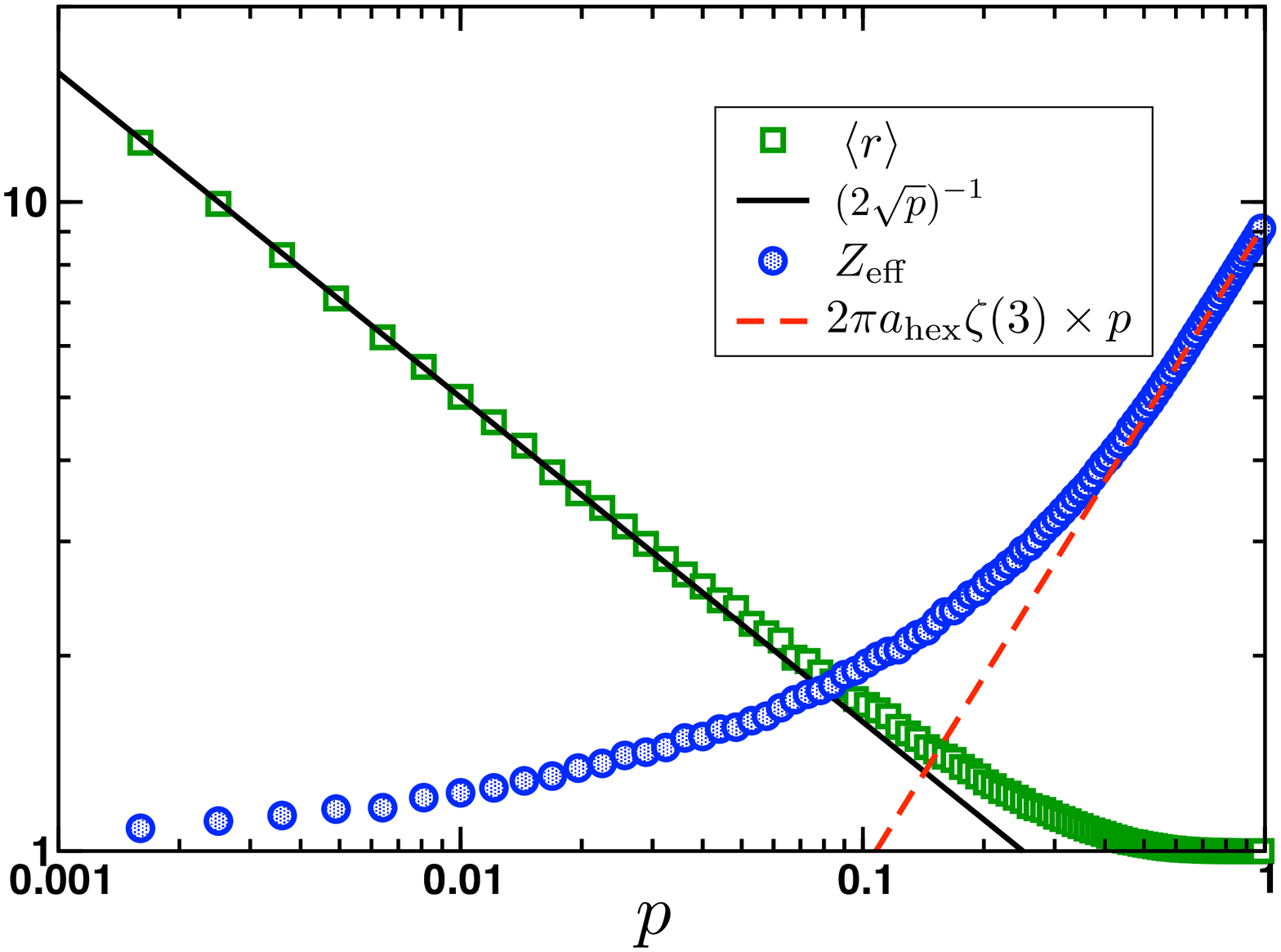}
\caption{{(Color online) 
Average distance $\langle r\rangle$ between nearest-neighbor magnetic moments, computed (green squares) on a $L=5000$ honeycomb lattice with a concentration $p$ of magnetic moments. The continuum formula $(2\sqrt{p})^{-1}$ (black line) holds at low doping. In this regime, the effective coordination number $Z_{\rm eff}$, computed over the same sample (blue circles), remains close to 1 and increase faster only for larger $p$, with the mean-field behavior (red dashed line) reached beyond $p\sim 0.25$.}}
\label{fig_avgr}
\end{center}
\end{figure}
}}

\end{document}